\documentclass{IET-Conf-Paper}
\usepackage{url}
\usepackage[inkscapelatex=false]{svg}
\usepackage{hyperref}
\usepackage{booktabs}
\usepackage{multirow}
\begin{document}

\title{Grid-Forming Control Methods for Weakly Connected Offshore WPPs}

\author{Sulav Ghimire\ad{1,2}\corr, Kanakesh V Kkuni\ad{1}, Simon C Jakobsen\ad{1,2}, Thyge Knueppel \ad{1}, Kim H Jensen\ad{1}, Emerson Guest\ad{1}, Tonny W Rasmussen\ad{2}, Guangya Yang\ad{2}}

\address{\add{1}{Siemens Gamesa Renewable Energy A/S, 7330 Brande, Denmark}
\add{2}{Department of Wind and Energy Systems, Technical University of Denmark, 2800 Kgs Lyngby, Denmark}
\email{sulav.ghimire@siemensgamesa.com}}


\begin{abstract}
Grid-forming control (GFC) has seen numerous technological advances in their control types, applications, and the multitude of services they provide. Some examples of the services they provide include black start, inertial frequency response, and islanded operation capabilities with the possibility of re-synchronization without the need of additional support from other devices such as storage. State of the art literature proposes a variety of GFCs which can provide single or multiple of these services. However, study of these different GFCs for weakly-connected offshore wind power plants (WPPs) based on time-domain simulation and focusing on the large signal disturbance is not well covered. This paper reviews some of the most researched grid-forming control methods applicable to offshore WPPs and provides a comparative investigation and discussion of their stability properties and applicability, especially when connected to a weak-grid. The paper also provides a discussion on the prerequisites and challenges surrounding the comparative study of different GFCs.
\end{abstract}

\maketitle

\section{Introduction}


Most of the converters used in grid-connected offshore WPPs are grid-following converters (GFLCs) which are designed to deliver power based on a reference point, and follow the grid frequency. They rely primarily on grid-voltage measurements from which a phase-locked loop (PLL) extracts grid phase and frequency information, thus synchronizing the converter with the grid. Performance of the rest of the cascaded controls (voltage and current control loops) depend on the PLL measurement of phase and frequency. These GFLCs find their alternatives in GFCs which use their voltage-source behavior to control the terminal voltage instead of behaving as a current-source and following the grid voltage, and provide frequency reference for the rest of the grid to follow instead of following the grid frequency based on PLL measurement. GFCs synchronize the converter to the grid via power loops without the need of PLLs (PLLs can impose several stability issues in weak grids which can lead to loss of synchronization \cite{ghimire2021dynamics, PLLInstAmir}.) GFCs can also emulate the behavior of the physical inertia of a synchronous machine by providing responses to system events such as increasing or reducing power during any load/generation changes or faults. They are also capable of running microgrids, or operate in islanding mode, and provide black-start to the grid.

Previous studies suggest that GFCs enable modern power systems to operate with little or no physical inertia by providing a synthetic inertia on top of the voltage and frequency reference points \cite{6200347}, however an inertial or power element like a DC bus or battery energy storage system (BESS) is still needed to support this synthetic inertia. GFCs are slowly being incorporated into the modern power grids with the rising need to integrate more converter-interfaced generation \cite{matevosyan2019grid, ndreko2018grid}. GFCs have become a thriving area of research due to their promising attributes and consequently, find extensive applications in microgrids \cite{marinakis2021grid, pogaku2007modeling}, converter-integrated power sources such as solar power plants \cite{pawar2021grid}, battery energy storage systems \cite{singh2015grid}, and WPPs \cite{roscoe2020response, jain2020grid}. It is anticipated that by the 2030s, they will be implemented in large island-grids, and by the 2050s, in bulk grids \cite{lin2020research}.
These long sea cables, combined with the desire to optimise the size of the WPP, often lead to low short circuit ratios (SCRs), which may challenge the WT stability in some cases. That is, in a weakly connected offshore WPP, the grid impedance is high, hence the back emf of the grid electrically distant, which, in severe cases, has been shown to challenge the stability \cite{7897172, 8450880}. Research and development in GFCs need to acknowledge their low inertia weak grid operations and study GFCs in all possible scenarios. Furthermore, offshore WPPs are generally connected to weak grids and stability is, with the growing unit size and penetration of WPPs in the generator mix, a major concern to the power grid and the power industry. GFCs show promising qualities to solve these issues \cite{dong2018small}, and thus encourages both academia and industry to pursue further research in GFCs. Following the growth of offshore WPPs, the inherent weak connections for offshore WPPs, and applicability of GFCs in offshore WPPs, there is clearly an incentive for more research and development in this area. This paper reviews different GFC methods and focuses on their performance comparison on offshore WPPs connected to a weak grid to determine the suitability of different control methods for such applications. This study can bolster the understanding of weak-grid issues and GFCs' performance on such weakly connected grids.

A short review of academic works and reviews on GFCs is provided hereby to illustrate the need of this paper. A review of 13 different types of GFCs classified under three categories: droop-based, synchronous machine based, and others (including virtual oscillator control (VOC), $H_2/H_\infty$ control, and frequency shaping control) is presented in \cite{9513281}. Detailed study on different capabilities of these control methods are discussed. Examples of a few real-life applications of GFCs in projects with BESS, gas-turbine black-start, and WPPs are also presented in the aforementioned paper. Reference \cite{9751441} presents the control structure of droop control, virtual synchronous machine (VSM), vitual oscillator control (VOC), and DC-link voltage control. Similarly, reference \cite{9759663} performs a comparative study of droop, VSM, VOC, and matching controls in terms of synchronization stability. Furthermore, \cite{9408354} shows the control structure of different power control loops, namely frequency droop control, power synchronization control, enhanced direct power control (EDPC), synchronverter, and synchronous power control (SPC.) It also illustrates the control structure of different inner loop control methods like droop control, PI-based voltage control, and PI-based reactive power control for SPC. Simulation and experimental results only for GFC with virtual admittance control and power synchronization control (PSC) are provided. Reference \cite{9627884} also gives an overview of different GFC methods like PSC, droop, low pass filter (LPF) droop, VSM, SPC, synchronverter, and VOC including their high voltage ride through (HVRT) and low voltage ride through (LVRT) capabilities. A review of pilot projects on grid-forming applications to BESS, HVDC stations, wind farms, power reserves, and other demonstrating projects is presented in \cite{musca2022grid} where grid-forming control facilitated ancillary services, black start capabilities, and islanded operation. Only one of the projects have provided specifics of the control methods. Several other studies have been performed on the review of grid-forming control algorithms \cite{9513281, 9751441, 9759663, 9408354, 9627884, musca2022grid}, however, none of them discussed about GFC applications or review for a weakly-connected offshore WPP case. Furthermore, none of these studies compared the operation of these GFCs via simulations or experiments; they only provide a general insight via the theoretical layout and characteristics of the converter control solely based on the available literature.

For GFCs to be applied for AC connected offshore WPPs -- which are often characterised by their weak grid connections (low SCR and low X/R ratio) -- it is crucial to have studies of different GFCs focused on such weak-grid connected scenario. Similarly, review works should properly address and investigate the applicability of their candidate GFCs on such weakly connected WPPs. The primary objective of this paper is to make a comparison of the dynamics of these control methods in terms of power, frequency, voltage, and currents following a disturbance in a weakly connected grid scenario and provide relevant analysis on the same. Tackling a common research gap on the aforementioned works, this paper studies different representative GFCs on weakly connected offshore WPPs and compares the dynamics of their operation in load addition and grid phase jump cases.

The key contributions of this paper are summarized as:
\begin{itemize}
    \item Literature review on different grid-forming control methods, and review papers overviewing and summarizing these methods; relevance of such review works on GFCs used in weakly-connected offshore WPPs.
    \item Modeling, simulation, and comparison of selected GFC candidates for weakly connected offshore WPPs.
\end{itemize}

The paper organization is as follows: section \ref{sec: Grid-Forming Control Methods} describes different GFC methods with a short literature review on each of them. Section \ref{sec: Methodology} shows the study framework, test system, and performance and evaluation criteria adopted for the paper. Discussion and analysis of the results are presented in section \ref{sec: Discussion} and the conclusion and an overview of possible research directions is presented in section \ref{sec: Conclusion}.


\section{Grid-Forming Control Methods}\label{sec: Grid-Forming Control Methods}

Five different GFC methods are chosen for this study. The control structure, mathematical representation of the control, an overview of the controls, and a brief literature review for each of these control methods are provided in this section. The control structure, its mathematical formulation, and overview are included to establish an understanding of the technicalities of the control methods, and the literature review is presented to understand the horizon covered by previous studies and whether or not there is any relevance to the specific case of weak-grid applications of the GFCs under study.

\subsection{Droop-based GFC}
\begin{figure}[htbp]
    \centering
    \includegraphics[width = 0.5\textwidth]{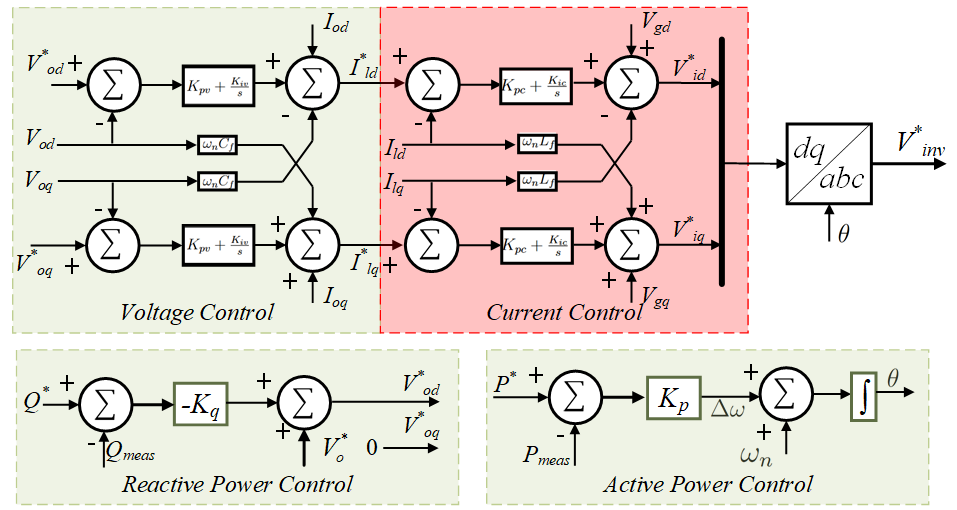}
    \caption{Control structure of droop-based GFC with inner loops.}
    \label{fig: droop-GFC Control}
\end{figure}

This GFC has droop-based active and reactive power control loops which also synchronizes the converter to the grid via power measurements. PI-based voltage and current controls with feed-forward decoupling loops are implemented in cascade with the power controller. A basic control diagram of the overall structure of droop-based GFC with inner loops is presented in figure \ref{fig: droop-GFC Control}.

The main dynamics of the power loop of this GFC can be written as follows:
\begin{eqnarray}
    \theta &=& \int[\omega_n+K_p(P^*-P_{meas})]dt,\\
    \label{eq: droop PQ loop 1}
    V_{od}^* &=& V_n^* - K_q(Q^*-Q_{meas}),\quad V_{oq}^* = 0.
    \label{eq: droop PQ loop 2}
\end{eqnarray}

The voltage control loop dynamics is governed by the following equations:
\begin{eqnarray}
    \label{eq: droop V loop 1}
    G_{PIv}(s) &=& K_{pv} + \frac{K_{iv}}{s},\\
    \label{eq: droop V loop 2}
    I_{ldq}^* &=& I_{odq} + G_{PIv}(s)(V_{odq}^*-V_{odq}) + j\omega_nC_fV_{odq}.
\end{eqnarray}

The current controller is similar to the voltage controller and it's dynamics is given as:
\begin{eqnarray}
    \label{eq: droop I loop 1}
    G_{PIc}(s) &=& K_{pc} + \frac{K_{ic}}{s},\\
    \label{eq: droop I loop 2}
    V_{idq}^* &=& V_{gdq} + G_{PIc}(s)(I_{odq}^*-I_{odq}) + j\omega_nL_fI_{ldq}.
\end{eqnarray}


Some droop-based control methods used in the literature involve a LPF before the implementation of the power-frequency droop control \cite{pogaku2007modeling}. This LPF takes away the fast reaction of a purely droop-based control and gives the GFC a VSM-like dynamics. Thus, the droop-based GFC candidate presented in this paper doesn't include a LPF before the power control loop to preserve the pure droop-nature of the control.

\subsection{Virtual Synchronous Machine based GFC}\label{sec: VSM}
\begin{figure}[htbp]
    \centering
    \includegraphics[width = 0.3\textwidth]{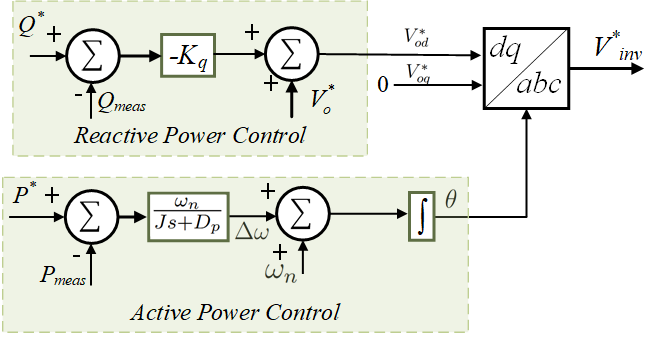}
    \caption{Control structure of VSM-based GFC without inner loops.}
    \label{fig: VSM-GFC Control outer}
\end{figure}

A VSM based GFC uses the emulated behavior of a synchronous machine to provide grid-forming behavior. Since it relies on the synchronous machine characteristics, the overall dynamics of this control is based on swing equation. A block diagram of VSM-based GFC without inner voltage and current loops is presented in figure \ref{fig: VSM-GFC Control outer}.

The dynamics of the outer loop of VSM is given as:
\begin{eqnarray}
    \label{eq: VSM PQ loop 1}
    \theta &=& \int\left[\omega_n+\frac{1}{Js+D_p}(P^*-P_{meas})\right]dt,\\
    \label{eq: VSM PQ loop 2}
    V_{od}^* &=& V_n^* - K_q(Q^*-Q_{meas}),\quad V_{oq}^* = 0.
\end{eqnarray}

\begin{figure}[htbp]
    \centering
    \includegraphics[width = 0.5\textwidth]{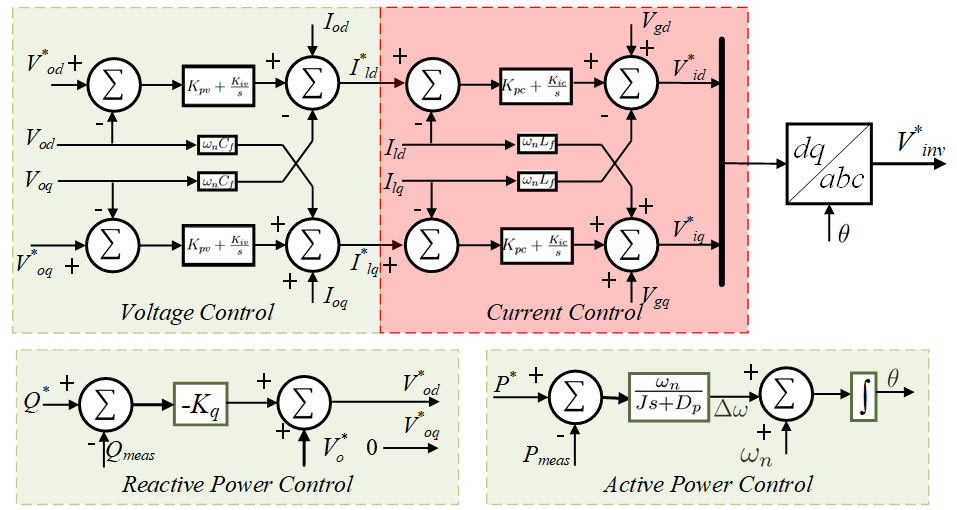}
    \caption{Control structure of VSM-based GFC with inner loops.}
    \label{fig: VSM-GFC Control}
\end{figure}

Faster inner control loops can be added to VSM-GFC to ensure fast response during transients. In this case, PI-based voltage and current control loops (equations \eqref{eq: droop V loop 1}--\eqref{eq: droop I loop 2}) are cascaded with the power control loops in the VSM model as shown in figure \ref{fig: VSM-GFC Control}.



A detailed implementation of VSM including the emulation of governor, prime mover, rotor dynamics, braking, and field excitation is presented in \cite{roscoe2016vsm} which is capable of inertia emulation with damping in addition to other characteristics like fault ride through (FRT), frequency ramping, and voltage step. Reference \cite{6891273} presents an automatic tuning method for VSM with cascaded inner loops similar to the VSM-GFC candidate presented in figure \ref{fig: VSM-GFC Control}. Here, the authors demonstrate conventional tuning approach for PI-based voltage and current controllers via pole cancellation and propose a new tuning approach via eigenvalue parametric sensitivity which could be beneficial for weak-grid operation of VSMs. Eight different levels of synchronous machine representation applicable for synchronous machine emulation to achieve grid-forming control is mentioned in \cite{chen2020modelling}, however modelling of only three such levels -- (a) swing equation based 2nd order model, (b) stator electromagnetic equation, governor, and AVR based 4th order model, and (c) symmetrical inductance matrices including stator electromagnetic transients based 7th order model -- are presented. The presented VSM method for DFIG application can support system frequency in a weak grid, but no insight is provided for a type-IV wind turbine. A detailed mathematical model of VSM implementation using a dynamic model based on electromagnetic equations is presented in \cite{7776920}. Furthermore, a closer insight on the effective damping and active and reactive power decoupling via virtual stator resistance is provided.

VSM is a promising method for GFC with prospective real-life applications to converter based generation. Since it follows some of the modelling and control approaches of a synchronous machines, and bypasses some of the physical constraints, it is considered as the GFC with the potential to run an entirely converter-based grid \cite{ierna2016effects}. Additionally, the plant hardware capabilities limit the support the converter is able to provide. Reference \cite{van2010virtual} presents two real life test cases for VSGs, one with $10\times5kW$ VSGs, and another with $1\times100kW$ VSG where the VSGs reacted to frequency changes by changing their power output and provided inertial response to improve the overall dynamic stability of the power grid.

\subsection{Virtual Admittance Control based GFC}\label{sec: Vadm}
\begin{figure}[htbp]
    \centering
    \includegraphics[width = 0.5\textwidth]{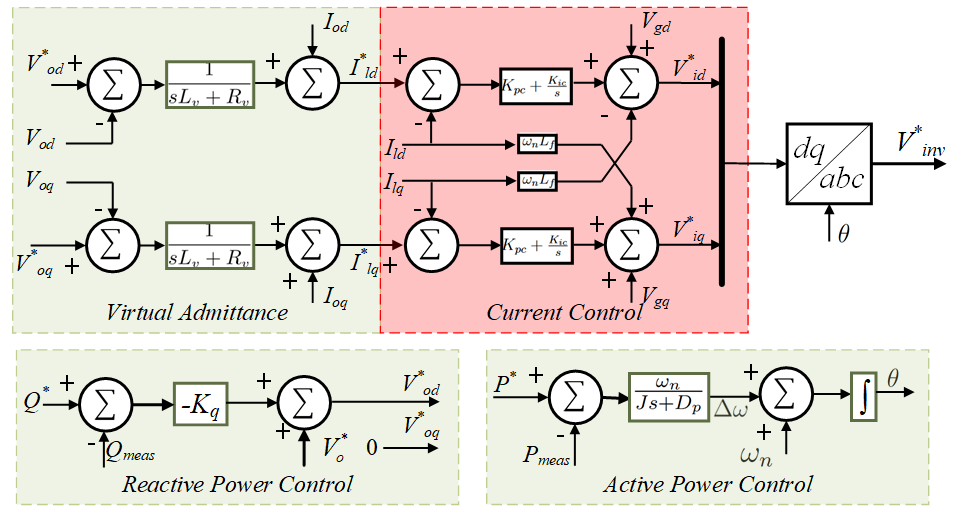}
    \caption{Control structure of virtual admittance-based GFC with inner current loop.}
    \label{fig: Vadm-GFC Control}
\end{figure}

This GFC method includes the application of a virtual admittance block to obtain expected current at the converter terminals based on the difference between the reference voltage and the measured voltage. This expected current is then fed to the current controller. The virtual admittance block doesn't control the converter voltage, but it defines the virtual impedance behind which the voltage-source lies via the virtual admittance block. The control diagram of this control method is illustrated in figure \ref{fig: Vadm-GFC Control}.

The dynamics of the power and current control loops in this control method is also the same as the VSM-based control, i.e. power loop is governed by equations \eqref{eq: VSM PQ loop 1} and \eqref{eq: VSM PQ loop 2}, and current controller is governed by equations \eqref{eq: droop I loop 1}--\eqref{eq: droop I loop 2}. The inner loop dynamics of the virtual admittance block is given as:
\begin{eqnarray}
    \label{eq: VAdm Yvirt 1}
    Y_{virt}(s) &=& \frac{1}{s L_v + R_v},\\
    \label{eq: VAdm Yvirt 2}
    I_{ldq}^* &=& I_{odq} + (V_{odq}^* - V_{odq})Y_{virt}(s).
\end{eqnarray}


Instead of implementing an inner voltage control loop, this method utilizes the virtual admittance block to estimate the current flowing at the converter terminal based on the measured voltage difference from the reference voltage \cite{6634621}. However, voltage control could be done in the outer loop voltage control either by an AVR or by a reactive power loop. Another use of virtual admittance block in converters is to minimize the circulating currents in a system with converters connected in parallel \cite{1490703}. Virtual admittance based method tackles issues related to the differentiation of the measured current which arises in virtual impedance based methods \cite{6634621}.

Synchronization issues may arise in weak grids while implementing virtual admittance control in rotating reference frame with hybrid synchronization control due to the cross-coupling of Q-f and P-V. Some researchers claim that this cross-coupling could be eliminated with the virtual admittance implementation in stationary reference frame \cite{10131594}, but since power is invariant to reference frame transformations, it is not clear how this cross-coupling could be eliminated. Furthermore, the measured quantities and reference variables in the stationary reference frames are sinusoids, which make it visually challenging to monitor and assess them.

\subsection{Proportional Resonant Control based GFC}\label{sec: PR Ctrl}
\begin{figure}[htbp]
    \centering
    \includegraphics[width = 0.5\textwidth]{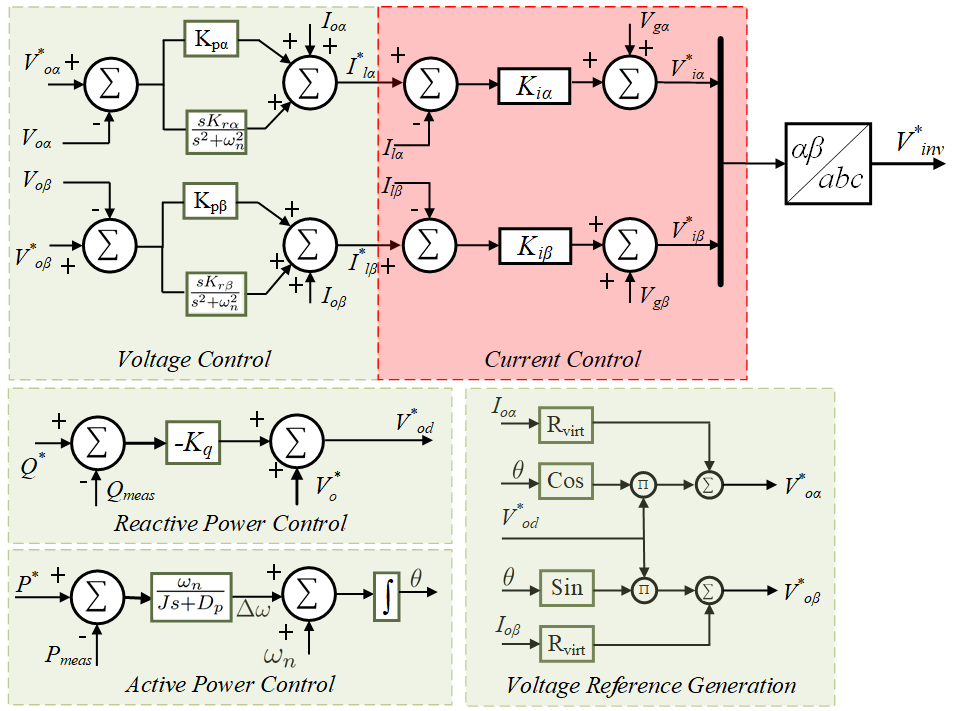}
    \caption{Control structure of GFC with proportional-resonant voltage controller.}
    \label{fig: PR-GFC Control}
\end{figure}

A proportional resonant (PR) control based GFC presented here includes a PR control based voltage controller and a proportional current controller in the inner loops. The outer control loops can be realized as a VSM (section \ref{sec: VSM}) and virtual admittance based GFC (section \ref{sec: Vadm}), i.e. a swing equation based active power loop and a droop based reactive power loop (equations \eqref{eq: VSM PQ loop 1} and \eqref{eq: VSM PQ loop 2}) with an additional option of AVR being cascaded between the reactive power loop and the voltage controller. A simple control diagram of a PR control based GFC with outer loops with swing equation based active power control and droop based reactive power control without AVR is presented in figure \ref{fig: PR-GFC Control}.

The voltage reference generation is governed by the following equations:
\begin{eqnarray}
    \label{eq: PRC Vref gen 1}
    V_{o\alpha\beta}^*  &=& R_{virt}I_{o\alpha\beta} + V_{od}^*\cdot e^{j\theta}.
\end{eqnarray}

The PR-control law for voltage control loops is given as:
\begin{eqnarray}
    \label{eq: PRC PR Vctrl 1}
    G_{PR\alpha\beta}(s) &=& K_{p\alpha\beta} +\frac{sK{r\alpha\beta}}{s^2+\omega_n^2},\\
    \label{eq: PRC PR Vctrl 2}
    I_{l\alpha\beta}^* &=& I_{o\alpha\beta} + (V_{o\alpha\beta}^*-V_{o\alpha\beta})G_{PR\alpha\beta}(s).
\end{eqnarray}

The current control is realized using a simple proportional-gain and is written mathematically as:
\begin{eqnarray}
    \label{eq: PRC ICtr 1}
    V_{i\alpha\beta}^* &=& V_{g\alpha\beta} + (I_{l\alpha\beta}^*-I_{l\alpha\beta})K_{i\alpha\beta}.
\end{eqnarray}

PR control is applied in stationary reference frame and leads to zero phase tracking and zero steady state error \cite{7052341}. A design method for PR voltage loop control is given in \cite{7446308} where authors provide methods to select the resonant gain for good steady-state and transient performance. Researchers \cite{9194347} also applied the method of dominant pole elimination (DPE) to design a PR-controlled GFC where the pole-zero pairs are independent of filter capacitance and inductance, thus making the DPE method insusceptible to filter parameter changes.

\subsection{Other Control Methods}
Virtual oscillator control (VOC) utilizes the nonlinear dynamics of a Van der Pol oscillator to attain grid-forming control \cite{6736685}. VOC inherently possesses a decentralized control strategy, uses local measurement to synchronize with the grid \cite{8722028} and ensures almost global asymptotic stability \cite{8638531, 8732453} which is guaranteed under specific bounds on the controller gains and references, and output power \cite{8732453}. Another work \cite{7776920} shows the study of a grid-forming converter with sliding-mode control (SMC) for inner-loop current control, and $H_2/H_\infty$ control for the outer-loop voltage control instead of the conventional PI controllers. The primary control via the outer loop has lower bandwidth, while the faster inner loop has higher bandwidth for momentary fluctuations. The $H_2/H_\infty$ control demonstrates robustness and optimal transient performance. It also provides continuous current reference signal, which otherwise would not have been possible through the SMC (since SMC creates discontinuous signals \cite{utkin2017sliding}). It is not explained well how the optimum control performance is obtained in the $H_2/H_\infty$ control, but the method itself needs a model-specific or system-specific optimization and makes the system difficult to be globalized. In comparison to the conventional nested-loop PI controller, the authors claim that their proposed method is more suitable for inaccurate model conditions and low switching frequencies.

\section{Methodology}\label{sec: Methodology}

\subsection{Network Model and SLD}\label{sec: Network Model and SLD}
\begin{figure}[htbp]
    \centering
    \includegraphics[width = 0.5\textwidth]{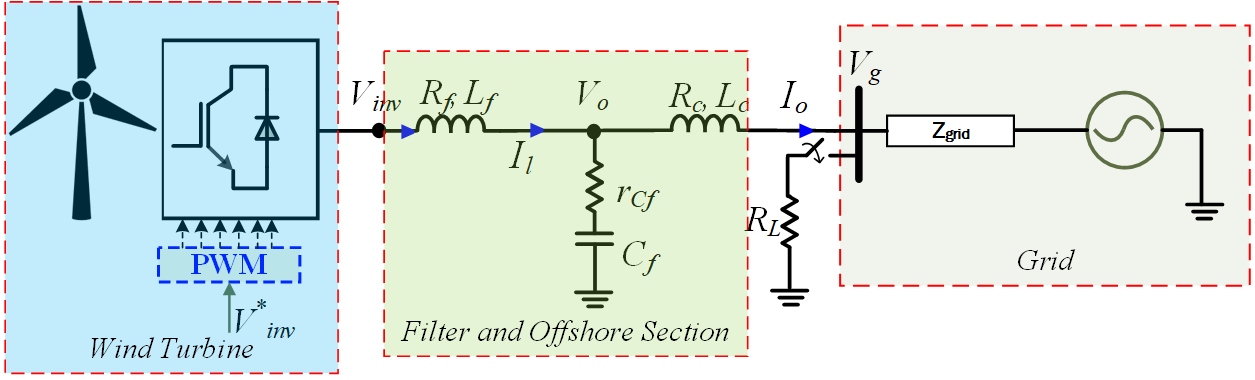}
    \caption{Single line diagram for the GF-control test setup used in the paper.}
    \label{fig: SLD}
\end{figure}

A full scale model of WPP is a complex system capable of representing both the generator side and network side controls, array cables, transformers, and sea cables, as well as the mechanical dynamics of the WT and its blades. However, in a WPP with homogeneous converter control and WTs, it is, sometimes, possible to aggregate them into a simplified benchmark model with single WT connected to the grid, when studying common mode stability of the WTs against the grid. A quasi-static benchmark test system model comprising of one WT (representing a WPP) connected to a weak-grid is developed for the studies. The single line diagram (SLD) of this system is presented in figure \ref{fig: SLD}. A switchable load ($R_L$) of power $P_L = 0.2\ pu$ is connected at the PCC for studies related to load change. A power grid is also connected at PCC with a grid impedance of $Z_{grid}$. A grid impedance of $Z_{grid} = (0.1178 + j0.5891)\ pu$ is used resulting in an $SCR$ of $\frac{1}{Z_{grid, pu}}=1.66$, and an $X/R$ ratio of $\frac{\Im(Z_{grid})}{\Re(Z_{grid})}=5$, which by today's standards would constitute a weakly connected WPP. This system serves as a test-bench for the presented control diagrams from section \ref{sec: Grid-Forming Control Methods}. $V^*_{inv}$ is the converter reference voltage fed to the converter PWM block by the respective control methods presented in section \ref{sec: Grid-Forming Control Methods}.


\subsection{Performance and Evaluation Criteria}\label{sec: Performance and Evaluation Criteria}
The relevant performance and evaluation criteria considered for the GFCs under studies are based on general GFC control system requirements. In order for a GFC to be suitable for application in grid-connected WPPs, they need to have (including but not limited to) the following characteristics following system events, for example load change, or grid phase shift.
\begin{itemize}
    \item Shallow frequency nadir and lower ROCOF
    \item Limited power, voltage, and current overshoots
    \item Fast response and smaller settling time
    \item No sustained oscillations following the event
\end{itemize}

\section{Discussion}\label{sec: Discussion}
All the converters undergo same test conditions, namely, $0.2\ pu$ load addition and $\pi/40\ rads$ grid phase shift, and the transient response of converters is studied. The test conditions are chosen such that they are not severe enough to trip the protection features of the system (for example any FRT mode or current prioritization modes) and facilitate the study of the control dynamic performances. The converters are tuned to match their steady-state behavior, i.e. all converter outputs are following the manually-set power reference of $P_{ref} = 1\ pu$, thus delivering $1\ pu$ of active power while maintaining a PCC voltage of $1\ pu$ and $1\ pu$ current. It must be noted that the frequency metrics here are outputs of power controllers and don't necessarily represent the grid/system's metrics as the grid is a stiff voltage source behind a Thevenin equivalent grid impedance, but represent the performance of the synchronizing power control loop. A brief account on the challenges to fulfill all prerequisites of an unbiased comparison of different GFCs is provided in section \ref{sec: Conclusion}.

\subsection{Load Change}
\begin{figure}[htbp]
    \centering
    \includegraphics[width =0.45\textwidth, trim = {1.8cm, 0cm, 1.8cm, 0cm}, clip]{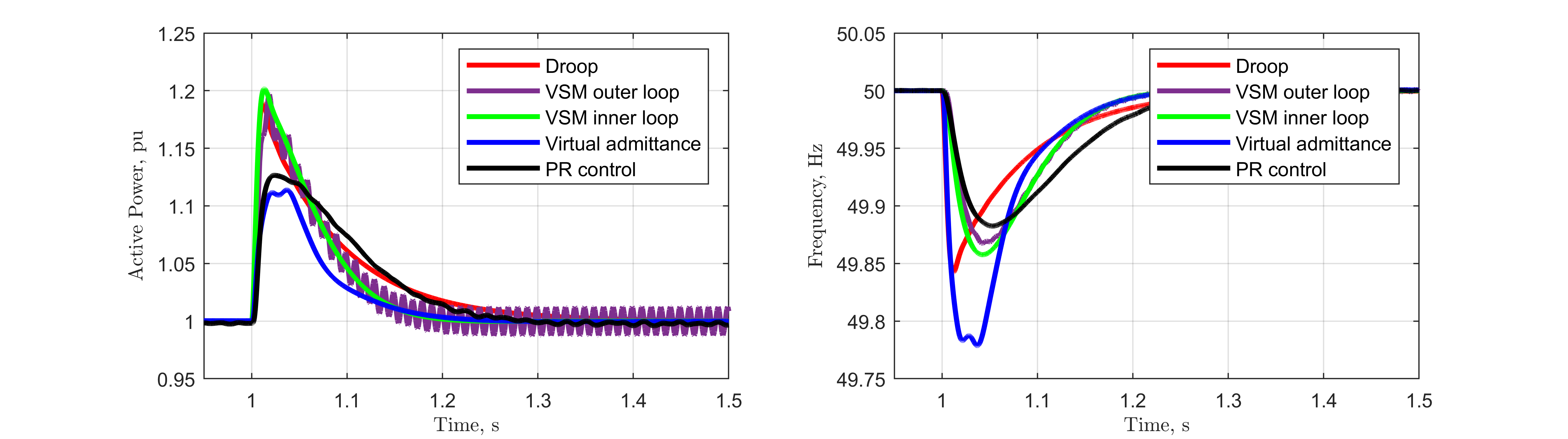}
    \caption{Active power and frequency comparison for different GFCs under $20\%$ load addition at $t=1s.$}
    \label{fig: Pf comparison GC}
\end{figure}

\begin{figure}[htbp]
    \centering
    \includegraphics[width =0.45\textwidth, trim = {1.5cm, 0cm, 1.8cm, 0cm}, clip]{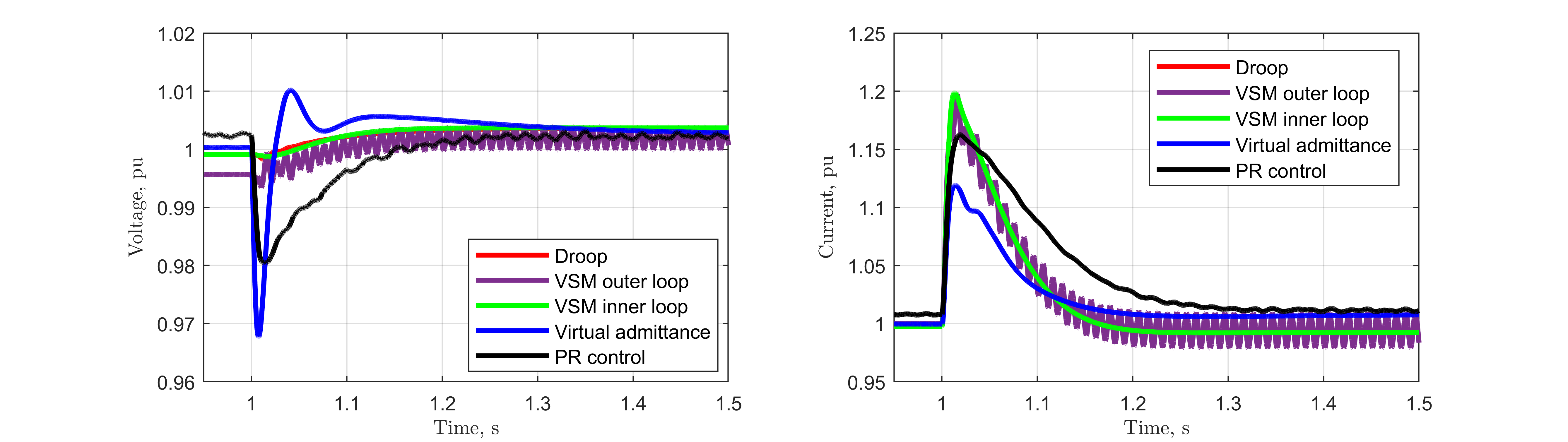}
    \caption{$d-$axis voltage and current comparison for different GFCs under $20\%$ load addition at $t=1s.$}
    \label{fig: VId comparison GC}
\end{figure}

The WT shown in figure \ref{fig: SLD} is equipped with different candidate GFC methods described in section \ref{sec: Grid-Forming Control Methods} and a load change of $0.2\ pu$ is introduced at $t = 1\ s$. The response in active power delivery, frequency, voltage, and currents were recorded and are presented in figures \ref{fig: Pf comparison GC} and \ref{fig: VId comparison GC}. Performance metrics like ROCOF, nadir, overshoots, and settling time of active power, frequency, voltage, and currents are summarized in table \ref{tab: Performance Evaulation LC}.

\begin{table*}[t]
	\centering
	\caption{Dynamic performance of grid-connected GFCs under a load addition of $0.2\ pu$ presented in terms of frequency, voltage, current, and power metrics.}
	\label{tab: Performance Evaulation LC} 
	\begin{tabular}{lcccccc}
		\hline
		  \multirow{2}{*}{GFC Method} & \multicolumn{3}{c}{Frequency (from power controller)} & \multicolumn{3}{c}{Overshoot and Settling Times} \\ \cmidrule(l){2-7} 
        & Nadir & ROCOF & Settling Time & Power         & Voltage         & Current        \\ \hline\hline
  		Droop              & 49.84 Hz & 13.47 Hz/s & 0.35 s            & 18.91 \%, 0.35 s            & -0.1726 \%, 0.1 s             & 19.87 \%, 0.2 s             \\
        VSM outer loop     & 49.87 Hz & 3.04 Hz/s  & \textgreater{}1 s & 19.69 \%, \textgreater{}1 s & -0.6426 \%, \textgreater{}1 s & 19.74 \%, \textgreater{}1 s \\
        VSM inner Loop     & 49.86 Hz & 3.33 Hz/s  & 0.25 s            & 20.16 \%, 0.25 s            & -0.1275 \%, 0.25 s            & 19.87 \%, 0.2 s             \\
        Virtual admittance & 49.78 Hz & 10.25 Hz/s & 0.25 s            & 11.37 \%, 0.25 s            & -3.2118 \%, 0.25 s            & 11.88 \%, 0.2 s             \\
        PR Control         & 49.88 Hz & 2.32 Hz/s  & 0.35 s            & 12.65 \%, 0.35 s            & -1.9379 \%, 0.35 s            & 16.25 \%, 0.35 s\\
		\hline 
	\end{tabular}
\end{table*}


Results show that although the initial response of droop based GFC following the disturbance is quite fast, its ROCOF is very high compared to other control methods. The fast response of droop-based GFC is attributed to its primary power control loop which provides a linear frequency response to any change in power with minimal control delay. It is also seen that the overall dynamic response trajectory of a VSM with and without the inner control loops are equivalent, except that the outer loop VSM has super-synchronous oscillations in its responses which sustains for a longer time (see figure \ref{fig: Pf comparison GC} and \ref{fig: VId comparison GC}.) Inclusion of inner loops in a VSM provides the system with better damping and thus faster settling times. Furthermore, the inertial behavior of VSM is clear in the results, i.e. in comparison to droop-based GFC, VSM provides a slower initial reaction to disturbances, however, the damping is better with short settling time and better frequency dynamics. Virtual admittance based method also exhibits a steep frequency changes and deeper nadir, although its settling time is low compared to other GFCs. The virtual admittance block of this control method is seen to have provided a damping on the power, which makes its damping property preferable to a VSM. The virtual admittance can define the voltage source behavior, thus shaping the power response following an event providing higher degrees of freedom in the control objective. PR control, on the other hand, has a longer settling time and sustained oscillations in its responses.




\subsection{Phase Shift}
\begin{figure}[htbp]
    \centering
    \includegraphics[width =0.45\textwidth, trim = {1.8cm, 0cm, 1.8cm, 0cm}, clip]{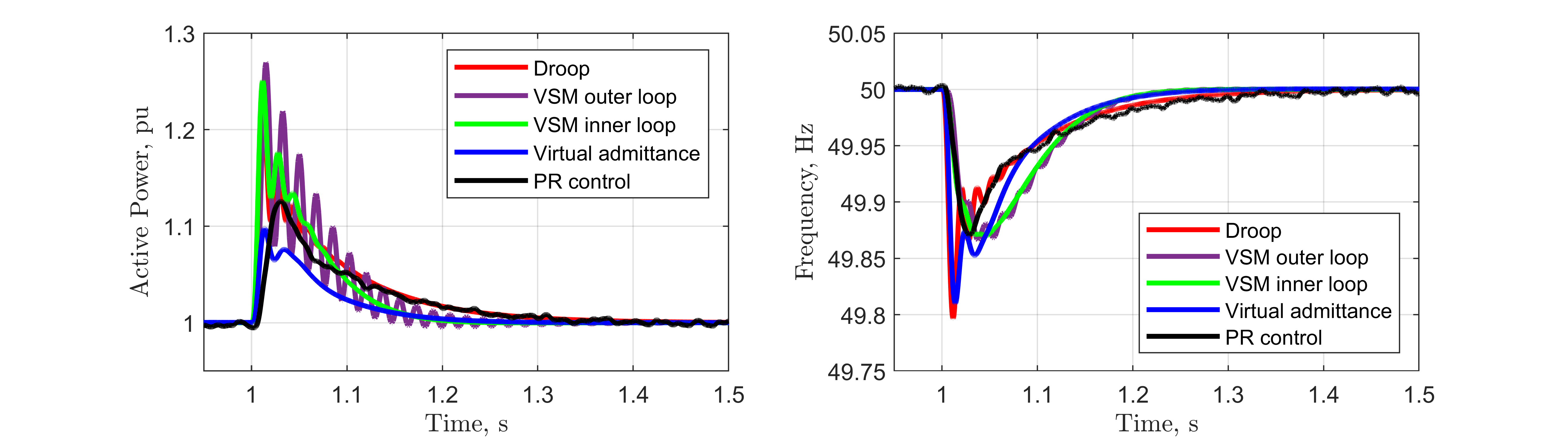}
    \caption{Active power and frequency comparison for different GFCs under $\pi/40\ rads$ grid phase shift at $t=1s.$}
    \label{fig: Pf comparison PShift GC}
\end{figure}

\begin{figure}[htbp]
    \centering
    \includegraphics[width =0.45\textwidth, trim = {1.5cm, 0cm, 1.8cm, 0cm}, clip]{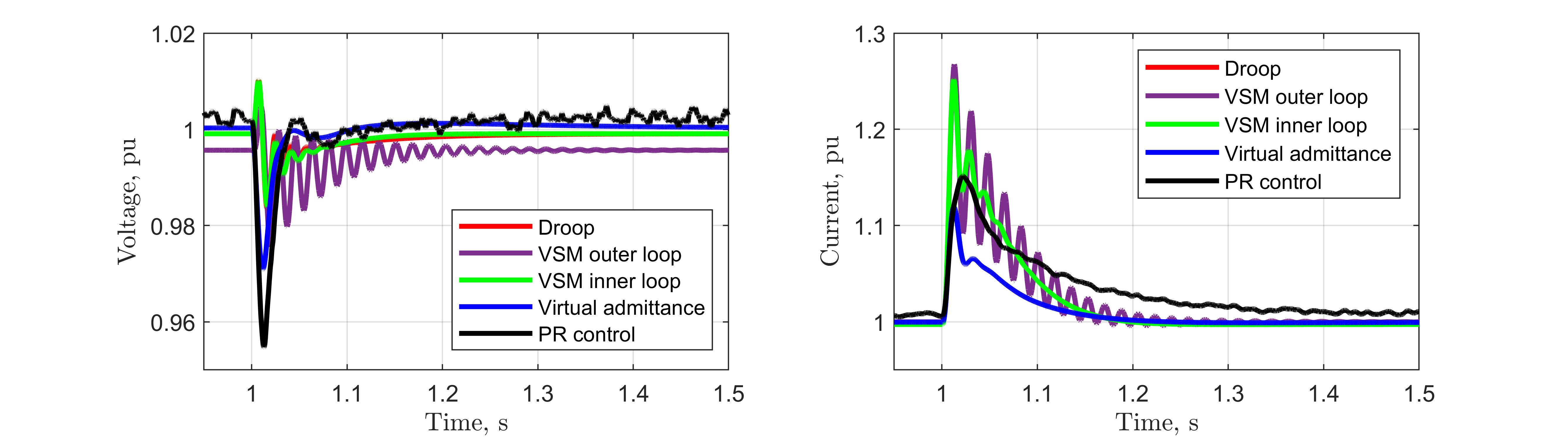}
    \caption{$d-$axis voltage and current comparison for different GFCs under $\pi/40\ rads$ grid phase shift at $t=1s.$}
    \label{fig: VId comparison PShift GC}
\end{figure}

A disturbance is introduced in the system as a grid phase shift of $\pi/40\ rads$ at time $t=1\ s$. The power and frequency responses of the respective GFCs is presented in figure \ref{fig: Pf comparison PShift GC} and the voltage and current responses are presented in figure \ref{fig: VId comparison PShift GC}. The obtained results are summarized in table \ref{tab: Performance Evaulation PS}.

A grid-phase shift leads to momentary misalignment of the converter and grid phases, in response to which the converter changes it's power output and frequency and attempts to re-align its phase with the grid. Results presented in figures \ref{fig: Pf comparison PShift GC} and \ref{fig: VId comparison PShift GC} show that a phase shift leads to under-damped oscillations in the system with droop, and VSM based controls. Virtual admittance based GFC reacts to the phase shift without sustaining any oscillations; the effect of virtual admittance based damping is apparent here. The synchronizing property of a PR-based control cited by many literature (refer to section \ref{sec: PR Ctrl}) is also seen here as the control returns back to synchronization without any oscillations. However, the settling time for PR control is still longer compared to other control methods. PR-control responds to the phase shift with a momentary sharp voltage dip with a relatively smaller current overshoot, however VSM (both inner and outer loops) have deeper current overshoots, which suggests that in the event of a more severe phase shift, VSM and droop-based controls are prone to current limit violations.

\begin{table*}[t]
	\centering
	\caption{Dynamic performance of grid-connected GFCs under a grid phase shift of $\pi/40\ rads$ presented in terms of frequency, voltage, current, and power metrics.}
	\label{tab: Performance Evaulation PS} 
	\begin{tabular}{lcccccc}
		\hline
		  \multirow{2}{*}{GFC Method} & \multicolumn{3}{c}{Frequency (from power controller)} & \multicolumn{3}{c}{Overshoot and Settling Times} \\ \cmidrule(l){2-7} 
        & Nadir & ROCOF & Settling Time & Power         & Voltage         & Current        \\ \hline\hline
  		Droop              & 49.80 Hz & 17.68 Hz/s & 0.4 s  & 24.39 \%, 0.5 s  & -1.65 \%, 0.1 s  & 25.19 \%, 0.25 s \\
        VSM outer loop     & 49.87 Hz & 3.56 Hz/s  & 0.5 s  & 26.95 \%, 0.5 s  & -2.44 \%, 0.35 s & 26.74 \%, 0.45 s \\
        VSM inner Loop     & 49.87 Hz & 3.80 Hz/s  & 0.3 s  & 25.04 \%, 0.25 s & -1.61 \%, 0.1 s  & 25.19 \%, 0.25 s \\
        Virtual admittance & 49.81 Hz & 13.77 Hz/s & 0.3 s  & 9.77 \%, 0.35 s  & -2.91 \%, 0.15 s & 11.84 \%, 0.25 s \\
        PR Control         & 49.87 Hz & 4.59 Hz/s  & 0.45 s & 11.25 \%, 0.4 s  & -4.53 \%, 0.15 s & 15.15 \%, 0.5 s \\ 
		\hline 
	\end{tabular}
\end{table*}




\section{Conclusion}\label{sec: Conclusion}
Five different GFC typologies were reviewed, simulated, studied, analyzed, and compared against each other. The study was performed on a benchmark test system representing a weakly connected offshore WPP under system events like load addition and grid phase shift. The analysis was performed keeping in mind the performance characteristics enlisted in section \ref{sec: Performance and Evaluation Criteria}.

Comparing different grid-forming control methods is a challenging task. The prerequisites for the comparison include not only their comparable steady-state behavior, but also the stability margins. However, matching one of the prerequisites can lead to a mismatch in others. Furthermore, the performance of these controls are highly susceptible to the controller tuning and tuning changes imply a different steady state behavior. Therefore, for the study performed in this paper, the study of dynamic transient behavior of the control was considered as the primary objective and thus a matching steady-state behavior of the controllers was considered as the prerequisite. Based solely on the presented results, VSM with inner loops and virtual admittance control show the most promising response to the applied disturbances in the setting of a weakly connected offshore WPP, however it must be noted that this is only valid for the considered prerequisites, control parameter tuning, and system condition(s.) Thus it is highly recommended to define the prerequisites and the comparison objective for further studies.

In addition to these simulations, small signal stability studies and eigenvalue analysis, time domain simulations, fault studies, and study for more severe phase shifts are necessary for a more detailed investigation and understanding. Other GFC methods like d/VOC, $H_1/H_\infty$, matching control, SMC, robust optimal control, could also be considered for comparison and analysis. Inclusion of these controls and further studies will be considered for future works.

\small
\section{Legal Disclaimer}\label{sec: Acknowledgements and Disclaimer}
This work was supported by Innovation Fund Denmark under the project Ref. no. 0153-00256B. Figures and values presented in this paper should not be used to judge the performance of Siemens Gamesa Renewable Energy technology as they are solely presented for demonstration purpose. Any opinions or analysis contained in this paper are the opinions of the authors and not necessarily the same as those of Siemens Gamesa Renewable Energy.


\section{References}
\bibliography{bibliography}
\bibliographystyle{ieeetr}

\end{document}